# Giant dielectric permittivity in Nb-doped rutile crystals

D. Nuzhnyy,[1,]* V. Bovtun,[1] J. Petzelt,[1] M. Savinov,[1] M. Kempa,[1] P. Levinský,[2] K. Beranová,[2] P. Vaněk,[1] T. Kmječ,[1] T. Ostapchuk,[1] P. Kužel,[1] J. Hlinka,[1] D. Crandles,[3] M. Cosco,[3] Y. Hashimoto,[4] H. Taniguchi,[4] S. Kamba[1]

*[1]Institute of Physics of the Czech Academy of Sciences, Na Slovance 2, 182 00 Prague 8, Czechia*
*[2]Institute of Physics of the Czech Academy of Sciences, Cukrovarnická 10/112, 162 00 Prague 6, Czechia*
*[3]Department of Physics, Brock University, St. Catharines, Ontario L2S 3A1, Canada*
*[4]Department of Physics, Nagoya University, Nagoya 464-8602, Japan*

## ABSTRACT

Dielectric properties of Nb-doped (~1.5 at.%) rutile single crystals were studied in the 10-300 K temperature range (at frequencies below the MHz range down to 0.3 K) in a broad frequency range, including terahertz and infrared range, to understand the origin of its giant permittivity. The results were fitted, modelled and compared with those of the undoped rutile crystal measured in the terahertz and infrared ranges. The primary effect originates from the near-electrode depletion layer of lower conductivity compared to the bulk (surface barrier-layer capacitor effect), which causes a strong thermally activated relaxation in the MHz dielectric spectra. In the higher frequency range, the main difference between doped and undoped crystals is the presence of an overdamped microwave excitation (central mode) in the doped crystal for both polarizations, persisting down to 10 K and not thermally activated. This accounts for the previously reported increased permittivity values down to 2 K – where all lower-frequency relaxations are frozen – compared to undoped crystals. It also explains why our low-frequency permittivity at 0.3 K exceeds its THz value. In similar Nb-doped rutile single crystals it was quite recently ascribed [V. A. Dinh et al., Phys. Rev. B **114**, 024112, (2026)] to microwave tunneling of molecular polarons. Polar phonons are affected only by a slightly increased damping.

## I. INTRODUCTION

The search for giant or colossal permittivity (CP) materials has been widespread since the beginning of the century, because of their potential applications as capacitors for energy storage [1,2]. For applications, small dielectric losses and high breakdown electric fields are required [3], therefore composite materials with conducting components, which frequently show CP, are not appropriate. The best studied single-phase dielectric material is the non-ferroelectric oxide $CaCu_3Ti_4O_{12}$, which shows a CP of ~$10^5$ up to the MHz range which is almost temperature independent in the 100 – 300 K range [4]. The CP was observed in ceramics as well as in single crystals and was first assigned to an intrinsic defect dipolar mechanism [4]. However, later the CP effect was shown to be caused mainly by a charge depletion layer (DL) effect on the sample-electrode interface called surface barrier-layer capacitor (SBLC) effect [5,6], and in ceramics also by the internal barrier-layer capacitor (IBLC) effect due to higher resistivity of the grain boundary layers compared to the grain bulk (Maxwell-Wagner effect). Since then, several other materials were revealed to show similar CP effects, as $La_{15/8}Sr_{1/8}NiO_4$ [7], Li+Ti co-doped NiO [8], $SrTiO_3$ ceramics sintered in nitrogen [9], doped or reduced $BaTiO_3$ nanoceramics [10,11]. The most attractive due to its low loss and small temperature dependence appeared to be the Nb + In co-doped rutile $TiO_2$ (NITO) ceramics [12-14].

Rutile is a well-known high-permittivity dielectric crystal with tetragonal structure (space group $P4_2/mnm$, room temperature permittivity $\varepsilon'_c \approx 170$ and $\varepsilon'_a \approx 86$) of a weakly incipient ferroelectric behaviour with a strong phonon mode slightly softening on cooling (soft mode – SM) polarized along the tetragonal *c*-axis [15-18]. For the *a*-axis polarization, the lowest-frequency phonon also weakly softens so that in the undoped rutile at liquid-He temperatures the low-frequency (LF) relative permittivities are $\varepsilon'_c \approx 257$ and $\varepsilon'_a \approx 111$ without appreciable dielectric dispersion below the polar-phonon range. In NITO ceramics, the dominant physical mechanism responsible for the CP effect was ascribed to the formation of localized electron-pinned defect-dipoles (EPDD) [12], which should produce a strong high-frequency (HF) relaxation in the grain bulk [19]. Later this interpretation was partially questioned, nevertheless the dielectric properties appeared to be quite attractive from the application point of view [20,21]. Understanding the above-mentioned interpretation would require a defect excitation (relaxation) presumably in the MHz-GHz range [19], which calls for the corresponding HF studies up to the microwave (MW), terahertz (THz) and infrared (IR) range, first performed for $CaCu_3Ti_4O_{12}$ [22] and related $Na_{1/2}Bi_{1/2}Cu_3Ti_4O_{12}$ ceramics [23]. More recently, in NITO ceramics it was shown that the CP effect is caused in a similar measure by both the SBLC and IBLC effects, without appreciable presence of the intrinsic EPDD relaxation in the HF-MW range [24-26]. Since then, also NITO single crystals were studied [27-29] in which the CP effect was even more pronounced (down to ~20 K the permittivity is > $10^5$ up to 1 MHz). Even down to 2 K, where the dispersion due to the charge freezing process was finished, the permittivity for $\boldsymbol{E} \parallel c$ remained several times higher than that in the undoped crystals, which was first assigned to the EPDD contribution [27,28]. However, such defects are also expected to be thermally activated so that their contribution to the dielectric response is expected to freeze out at sufficiently low temperatures. Later it was revealed that appropriate annealing of NITO crystals results in radical reduction of the CP effect with vanishing dispersion below the MHz range [30]. Still, the permittivity values remained appreciably higher than those in the undoped rutile crystal and were assigned to the EPDD even if no dielectric data above the MHz range are so far available. Recent studies in NITO and in other similarly co-doped rutile crystals revealed [31] that the CP effect can be eliminated by appropriate annealing together with the $In_2O_3$ powder which introduces the missing $In^{3+}$ ion concentration into the sample and strongly reduces the losses, compensating for the charges introduced by the aliovalent $Nb^{5+}$ doping. Also, just Nb-doped (1 at.%) rutile crystal (NTO) [31] was shown to display a very similar CP effect as that in NITO. However, the reason why the plain $Nb^{5+}$ doping also provides the CP effect was not discussed until very recently [29], where the increased permittivity was ascribed to a contribution of small polarons at neighboring $Nb^{5+}$-$Ti^{3+}$ ions.

* corresponding author: nuzhnyj@fzu.cz

In this paper we report the first dielectric studies of NTO crystals in a very broad frequency range including LF, HF, MW, THz and IR spectroscopy and in a broad temperature range from 300 K down to 10 K in the THz and IR range and down to 0.3 K at LF to discuss its CP and DC and AC conductivity properties.

## II. EXPERIMENTAL

The (001) and (110) plane-parallel plates for our studies were cut from NITO crystal grown in an optical floating zone furnace [28]. Subsequently, after finishing our dielectric measurements, we checked the dopant concentrations of our samples by X-ray fluorescence (XRF) which revealed that the In-concentration was below the detection limit (100 ppm). Therefore, we will further denote this crystal as NTO and treat it as a plain Nb-doped rutile crystal. The negligible In-concentration in another NITO crystal was recently also indicated by XRF in Ref. [31].

In our XRF experiment, the Orbis® PC Micro-XRF spectrometer (EDAX, Ametek) with 50 kV Rh X-ray tube, poly-capillary optics, diameter of the incident beam 30 µm, silicon-drift detector, FWHM resolution ≈130 eV/Mn, was used for checking the elements compositions. The measurements under vacuum and not covered sample allow for the detection of all elements from Na up to U. Non-standard fundamental parameter method in Vision software [32] was used for semiquantitative approximation with the accuracy of ~0.5 wt.% for the main component and somewhat lower for the trace (<5%) elements.

X-ray photoelectron spectroscopy (XPS) measurements were performed using the AXIS Supra photoelectron spectrometer (Kratos Analytical). More details on the XPS measurement and analysis are provided in Supplementary Material.

DC conductivity between 300 and 2 K was measured using the 4-point method with a Keithley 2401 Source Meter sourcing the current and Keithley 2001 Multimeter measuring the voltage. Silver paint was used to affix four silver wires to the sample with the dimensions 3.3×1.3×0.8 mm$^3$. Data in the temperature range of 300 – 10 K were collected in a closed-cycle helium cryostat, while in the range of 20 – 2 K a Quantum Design Physical Property Measurement System (QD PPMS) was used to cool the sample.

LF complex dielectric permittivity at 1 Hz – 1 MHz frequencies and 0.3 – 300 K temperature range was measured using a Novocontrol Alpha-AN High-Performance Frequency Analyzer in conjunction with the cryostat utilizing a single-shot $^3$He insert with embedded coaxial cables. The temperature rate and the applied AC electric field was about 2 K/min and (2.5 – 1) V/mm, respectively. The specimens were fabricated as thin plane-parallel polished plates of 0.4 - 1 mm thicknesses. Ag or Au electrodes were used. Ag electrodes were formed by painting surfaces with the Pelco colloidal silver. Au electrodes were evaporated using a Bal-Tex SCD 050 sputter coater onto the principal faces of the plates. The contacts for applying the electric field were provided by silver wires fixed to the electrodes by a silver paste.

Dielectric measurements in the HF range of 1 kHz – 1 GHz were carried out using a Keysight E5061B network analyzer, a Novocontrol BDS 2100 coaxial sample cell and a Cold Edge CH-204N closed-cycle He cryostat on heating (temperature range 10 – 305 K, heating rate 0.5 K/min). The samples were cylinders of 6.0- and 1.3 mm height and 1.0- and 1.2 mm diameter for the $\boldsymbol{E} \parallel c$ and $\boldsymbol{E} \perp c$ measurements, respectively, and sputtered Au electrodes on the flat surfaces. Complex permittivity was numerically calculated from the measured complex impedance, taking into account the electromagnetic field distribution as well as the geometry of both the sample and measurement cell.

Complex MW response at 5.8 GHz was measured using the composite dielectric resonator method [33]. The $TE_{01\delta}$ resonance frequency, quality factor and insertion loss of the base cylindrical dielectric resonator with and without the sample were recorded during the heating from 10 to 400 K with a temperature rate of 0.5 K/min in a Janis CCS-400T/204 closed-cycle He cryostat. The same samples without electrodes as for the THz measurements ((110) and (001) plates of 100 and 255 μm thickness, respectively) were placed on top of the base dielectric resonator. The resonator was measured in the cylindrical shielding cavity using the transmission setup with a weak coupling by an Agilent E8364B network analyzer. MW complex permittivity of the sample was calculated from the acquired resonance frequencies and quality factors of the base and composite resonators. For the (110) sample, due to the circular distribution of the in-plane electric field component, only the average of $\varepsilon_c^*$ and $\varepsilon_a^*$ could be determined. Due to the significant difference in thickness between the base resonator (3.8 mm) and samples, the estimate of the MW complex permittivity value is rather crude. However, its relative temperature dependence is reliable. Therefore, we have normalized the calculated room-temperature data to the more accurate HF and THz data and present mainly their temperature dependences.

In the THz spectral range of 200 GHz – 2 THz, a custom-made time-domain THz spectrometer based on Ti:sapphire femtosecond laser [34] was used in combination with an Oxford Instruments Optistat continuous He-flow cryostat with mylar windows to obtain the complex dielectric response for both polarizations ($\boldsymbol{E} \parallel c$ and $\boldsymbol{E} \perp c$) and the temperature range of 5 – 300 K. Measurements were performed in the transmission configuration (measuring of the transmission signal and its phase) with the polished plane-parallel NTO plates of (110) and (001) orientations of 100 and 255 μm thickness, respectively.

Polarized IR reflectivity measurements at near normal incidence on polished sample surfaces were conducted from 80 to 5500 cm$^{-1}$ using a homemade reflectometer attached to a Bruker IFS 113v Fourier Transform Infrared spectrometer with polarizer/beamsplitter/detector combinations appropriate for the far/mid IR ranges. The *in-situ* gold evaporation technique [35] was employed to normalize the data. The samples were mounted on a Janis continuous-flow He cryostat permitting measurements to be made between 4.2 and 300 K.

## III. RESULTS

### A. Sample characterization

Three samples from the NITO boule were characterized using XRF at 17, 7, and 13 points, respectively. An additional sample was analyzed on the surface (at 9 points) and on the fracture surface (at 9 points). The In concentration was below the detection limit in all samples. The Nb concentration was ~1.5 ± 0.2 at.% (~2.9 ± 0.4 wt.%). The absolute error might be higher, at low element concentrations, up to relative 50% when the fundamental parameter method was used, since the standard sample was not accessible for this composition.

The XPS results confirmed that titanium is present in the rutile structure [36,37] with the $Ti/O_{Lattice}$ ratio of 0.53. Approximately 3.2 at.% of titanium is in the $Ti^{3+}$ oxidation state. Niobium accounts for approximately 1.6 at.% of the cations in the oxide structure and appears in a single chemical state originating from Nb substituting Ti in $TiO_2$ [38,39]. Carbon, together with traces of K, Ca and Zn, was detected by XPS. These elements are attributed to surface contamination. In addition, K, Ca and Zn were not detected in the bulk by XRF, further supporting their assignment to surface contamination.

### B. Broadband dielectric spectra

Let us summarize the results of LF, HF, MW, THz and IR experiments and analyze the dielectric response of NTO up to ~1 THz in the 10-300 K temperature range. The LF and HF data were obtained on samples with Au electrodes, whereas the higher-frequency measurements were performed on samples without electrodes. Real and imaginary parts of the complex permittivity $\varepsilon^*(f) = \varepsilon'(f) - i\varepsilon''(f)$ and AC conductivity $\sigma(f) = 2\pi f \varepsilon_0 \varepsilon''(f)$ spectra are used for the analysis. The $\varepsilon'(f)$, $\varepsilon''(f)$ and $\sigma(f)$ spectra for both $\boldsymbol{E} \parallel c$ and $\boldsymbol{E} \perp c$ polarizations are shown in Fig. 1. Several ranges of dielectric dispersion are observed. Up to the THz range the spectra can be well fitted to Eq. 1, which accounts for the plateau in the LF conductivity by implementing the Drude model (with the high damping of $\Gamma$=100 THz which yields a small essentially frequency independent DC conductivity contribution $\sigma_0$ up to the IR range) and relaxation processes described by the Cole-Cole model:

$$\varepsilon^*(f) = \varepsilon^*_{IR}(f) + \sum_j \frac{\Delta\varepsilon_{Rj}}{1 + \left(\frac{if}{f_{Rj}}\right)^{1-\alpha_j}} + \frac{\sigma_0}{\varepsilon_0 f(i - f/\Gamma)} \tag{1}$$

where $\Delta\varepsilon_{Rj}$ is the dielectric strength of the $j$-th Cole-Cole relaxation, $f_{Rj}$ is the corresponding relaxation frequency (matching frequency of its dielectric loss spectrum maximum) and $\alpha_j$ varies between 0 and 1 from the Debye relaxation to a frequency-independent losses; $\varepsilon_0$ is the permittivity of vacuum and $\varepsilon^*_{IR}(f)$ is the THz-IR response evaluated from the independent fits to IR reflectivity and THz-MW data using the factorized formula of damped harmonic oscillators [40], which includes contributions from an overdamped central mode in the MW range (CM) and polar phonons:

$$\varepsilon^*_{IR}(f) = \varepsilon_\infty \prod_{j=1}^{n} \frac{f^2_{LOj} - f^2 + if\gamma_{LOj}}{f^2_{TOj} - f^2 + if\gamma_{TOj}}, \quad R(f) = \left| \frac{\sqrt{\varepsilon^*_{IR}(f)} - 1}{\sqrt{\varepsilon^*_{IR}(f)} + 1} \right|^2, \tag{2}$$

where $\varepsilon_\infty$ is the permittivity due to electronic processes far above the optical phonon frequencies, obtained from mid-infrared reflectivity fits, $f_{TOj}, f_{LOj}$ are transverse and longitudinal frequencies of the $j$-th polar phonon, respectively, and $\gamma_{TOj}, \gamma_{LOj}$ are the corresponding damping constants. Broadband fits using Eq. 1 can be used up to the THz range; at higher frequencies they are not used since the Cole-Cole relaxations, which do not obey the conductivity sum rule [41], contribute unphysically nonzero to the dielectric losses up to infinite frequencies. Therefore, for the fits in the THz-IR range we use only Eq. 2. The fitting curves presented in the figures were obtained by merging the respective fits at around 7 THz. All the fit parameters of the $\boldsymbol{E} \parallel c$ and $\boldsymbol{E} \perp c$ spectra are listed in Tables I and II, respectively, along with the dielectric strengths $\Delta\varepsilon_j$ of the corresponding THz-IR excitations:

$$\Delta\varepsilon_j = \frac{\varepsilon_\infty}{f_{TOj}^2}\frac{\prod_k\left(f_{LOk}^2 - f_{TOj}^2\right)}{\prod_{k\neq j}\left(f_{TOk}^2 - f_{TOj}^2\right)}. \qquad (3)$$

The LF $\sigma(f)$ plateau values correspond to the effective DC conductivity (Fig. 1c,f) fitted by the Drude model. Another $\sigma(f)$ plateau is seen in the MHz frequency range. Two Cole-Cole relaxations (R1 and R2) were fitted at frequencies below 1 GHz. They are overlapping and R1 is partially screened by the Drude conductivity. An overdamped CM in the 10 GHz range is required to join the HF and THz spectra, even if it does not fit well the measured data at 5.8 GHz, whose absolute values are of limited reliability. We emphasize that the HF and THz data are quite accurate (cca +/-5%) and, below ~100 K, cannot be connected without an additional MW excitation (CM).

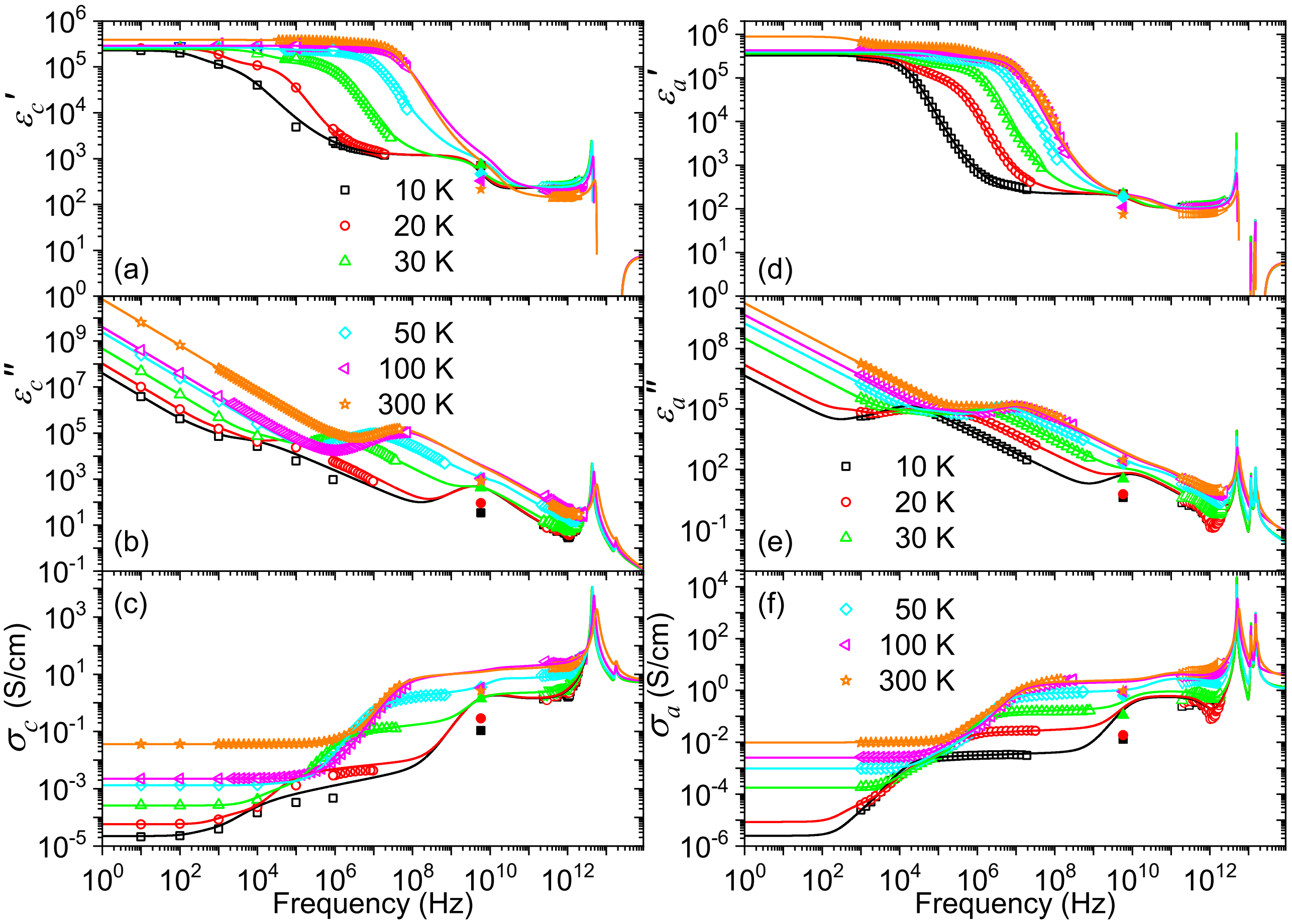

Fig. 1. (a,b,c) Broadband dielectric spectra for $\boldsymbol{E} \parallel c$ and (d,e,f) $\boldsymbol{E} \perp c$ polarizations: (a,d) permittivity, (b,e) dielectric loss, (c,f) conductivity at selected temperatures. Experimental data are presented as symbols; lines correspond to the fits by Eqs. 1 and 2.

Table I. Parameters of the Cole-Cole relaxations and of the CM and SM oscillators from the broadband fits (Eqs. 1-3) of the dielectric spectra for $\boldsymbol{E} \parallel c$. The DC conductivities $\sigma_0$ of the Drude term are 2.2×10$^{-5}$, 2.26×10$^{-3}$ and 3.66×10$^{-2}$ S/cm for 10, 100 and 300 K, respectively. * and ** denote the CM and SM, respectively.

| Temperature | Cole-Cole relaxation | | | Generalized oscillator ($\varepsilon_\infty = 8$) | | | | |
|---|---|---|---|---|---|---|---|---|
| | $f_{Rj}$ (Hz) | $\alpha_j$ | $\Delta\varepsilon_{Rj}$ | $f_{TOj}$ (Hz) | $\gamma_{TOj}$ (Hz) | $f_{LOj}$ (Hz) | $\gamma_{LOj}$ (Hz) | $\Delta\varepsilon_j$ |
| 10 K | 231 | 0.09 | 1.06×10$^5$ | *1.09×10$^{10}$ | 3.29×10$^{10}$ | 2.37×10$^{10}$ | 4.34×10$^{10}$ | 937.3 |
| | 4.85×10$^3$ | 0.27 | 1.23×10$^5$ | **4.36×10$^{12}$ | 2.12×10$^{11}$ | 1.68×10$^{13}$ | 1.96×10$^{12}$ | 224.9 |
| | | | | 1.69×10$^{13}$ | 2.2×10$^{12}$ | 2.36×10$^{13}$ | 1.59×10$^{12}$ | 0.1 |
| 100 K | 2.77×10$^5$ | 0.1 | 1.43×10$^4$ | *1.84×10$^{10}$ | 3.77×10$^{10}$ | 3.51×10$^{10}$ | 5.83×10$^{10}$ | 510.4 |
| | 4.93×10$^7$ | 0.11 | 2.76×10$^5$ | **4.79×10$^{12}$ | 4.28×10$^{11}$ | 1.72×10$^{13}$ | 1.35×10$^{12}$ | 186.5 |
| | | | | 1.73×10$^{13}$ | 1.48×10$^{12}$ | 2.37×10$^{13}$ | 1.84×10$^{12}$ | 0.1 |
| 300 K | 5.42×10$^5$ | 0.3 | 6.28×10$^4$ | *1.84×10$^{10}$ | 3.77×10$^{10}$ | 3.51×10$^{10}$ | 5.83×10$^{10}$ | 351.5 |
| | 4.47×10$^7$ | 0.07 | 3.27×10$^5$ | **5.64×10$^{12}$ | 1.2×10$^{12}$ | 1.76×10$^{13}$ | 2.65×10$^{12}$ | 126.1 |
| | | | | 1.78×10$^{13}$ | 2.83×10$^{12}$ | 2.39×10$^{13}$ | 2.85×10$^{12}$ | 0.2 |

Table II. Parameters of the Cole-Cole relaxations and generalized oscillator parameters from the broadband fits (Eqs. 1-3) of the dielectric spectra for $\boldsymbol{E} \perp c$. The DC conductivities $\sigma_0$ of the Drude term are 2.5×10$^{-6}$, 2.58×10$^{-3}$ and 9.9×10$^{-3}$ S/cm for 10, 100 and 300 K, respectively. * and ** denote the CM and SM, respectively.

| Temperature | Cole-Cole relaxation | | | Generalized oscillator ($\varepsilon_\infty = 6.1$) | | | | |
|---|---|---|---|---|---|---|---|---|
| | $f_{Rj}$ (Hz) | $\alpha_j$ | $\Delta\varepsilon_{Rj}$ | $f_{TOj}$ (Hz) | $\gamma_{TOj}$ (Hz) | $f_{LOj}$ (Hz) | $\gamma_{LOj}$ (Hz) | $\Delta\varepsilon_j$ |
| 10 K | 1×10$^3$ | 0 | 3.46×10$^4$ | *8.05×10$^{10}$ | 7.39×10$^{11}$ | 1.16×10$^{11}$ | 7.36×10$^{11}$ | 113.8 |
| | 1.58×10$^4$ | 0.05 | 2.92×10$^5$ | **4.95×10$^{12}$ | 4.01×10$^{10}$ | 1.1×10$^{13}$ | 9.93×10$^{10}$ | 97.4 |
| | | | | 1.15×10$^{13}$ | 1.92×10$^{11}$ | 1.34×10$^{13}$ | 2.23×10$^{11}$ | 1 |
| | | | | 1.5×10$^{13}$ | 3.16×10$^{11}$ | 1.69×10$^{13}$ | 8.44×10$^{11}$ | 2.5 |
| | | | | 1.69×10$^{13}$ | 9.14×10$^{11}$ | 2.45×10$^{13}$ | 6.9×10$^{11}$ | 0.1 |
| 100 K | 2.27×10$^5$ | 0.15 | 1.5×10$^5$ | *1.34×10$^{11}$ | 4.81×10$^{11}$ | 1.79×10$^{11}$ | 4.92×10$^{11}$ | 80.1 |
| | 1.06×10$^7$ | 0.04 | 2.78×10$^5$ | **5.13×10$^{12}$ | 3.8×10$^{11}$ | 1.1×10$^{13}$ | 1.5×10$^{11}$ | 92.1 |
| | | | | 1.15×10$^{13}$ | 2.51×10$^{11}$ | 1.34×10$^{13}$ | 3.69×10$^{11}$ | 1 |
| | | | | 1.52×10$^{13}$ | 4.22×10$^{11}$ | 1.76×10$^{13}$ | 2.07×10$^{12}$ | 2.7 |
| | | | | 1.77×10$^{13}$ | 1.78×10$^{12}$ | 2.45×10$^{13}$ | 1.45×10$^{12}$ | 0.1 |
| 300 K | 958 | 0.15 | 3.99×10$^5$ | *2.17×10$^{11}$ | 1.01×10$^{11}$ | 3.05×10$^{11}$ | 1.01×10$^{12}$ | 80.3 |
| | 6.46×10$^5$ | 0.01 | 1.9×10$^5$ | **5.62×10$^{12}$ | 8.99×10$^{11}$ | 1.09×10$^{13}$ | 3.85×10$^{11}$ | 73 |
| | 1.14×10$^7$ | 0.04 | 3.02×10$^5$ | 1.14×10$^{13}$ | 6.69×10$^{11}$ | 1.33×10$^{13}$ | 8.94×10$^{11}$ | 1.1 |
| | | | | 1.51×10$^{13}$ | 9.53×10$^{11}$ | 1.7×10$^{13}$ | 1.59×10$^{12}$ | 2.6 |
| | | | | 1.72×10$^{13}$ | 1.54×10$^{12}$ | 2.46×10$^{13}$ | 2.22×10$^{12}$ | 0.3 |

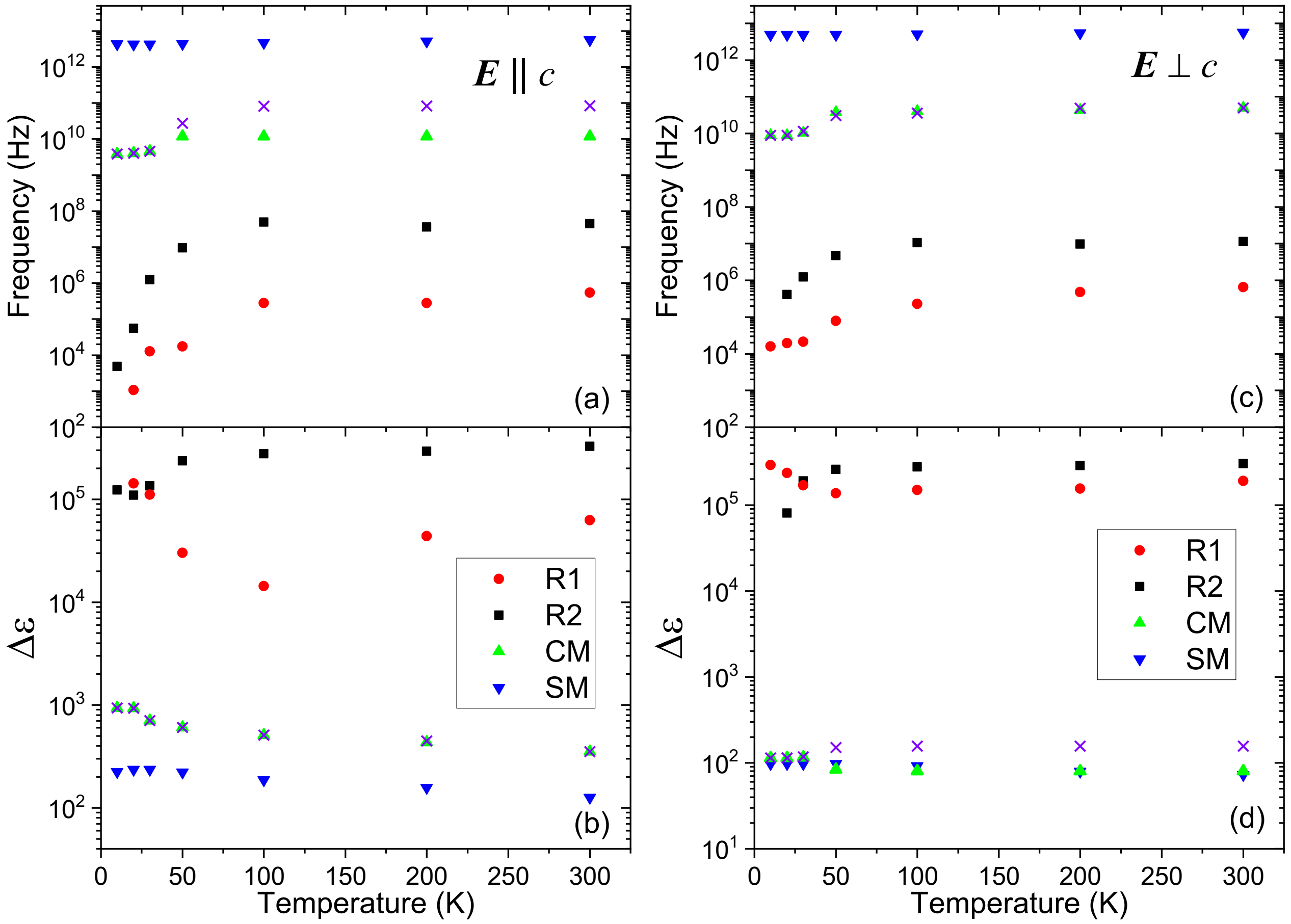

Fig. 2. (a,c) Temperature dependence of the fitted main transverse excitation frequencies in NTO up to the SM. For the overdamped CM, the maximum-loss frequency is shown. (b,d) Corresponding dielectric strengths for both polarizations. Crosses represent CM parameters obtained from fits to the THz–IR spectra alone using Eq. 2.

In Fig. 2 we plot the temperature dependences of the fitted excitation frequencies up to the IR range including the SM and their dielectric strengths from our fits. The higher-frequency polar phonon frequencies (not shown in Fig. 2) do not differ appreciably from that of the undoped rutile (see later in Discussion) and their temperature dependence is small. One can see that both Cole-Cole relaxations are thermally activated with dielectric strengths in the order of $\sim 10^5$, but the higher-frequency R2 is mostly stronger. The overdamped CM in the 10-100 GHz range is not thermally activated. As shown in Fig. 2, its characteristic frequency at higher temperatures depends on the fitting approach, because Cole-Cole relaxations included in the fits using Eq. 1 contribute unphysically to the losses even at much higher frequencies. Therefore, for the CM frequencies and strengths, we prefer to use the fits by Eq. 2, which yield slightly higher frequency and dielectric strengths values.

In Fig. 3 we plot the temperature dependence of the permittivity and loss at selected frequencies up to the THz range. It is seen that the dielectric permittivity is nearly temperature independent above 100 K for both polarizations below ~3 MHz (Fig. 3a,c). Diffuse $\varepsilon''(T)$ maxima observed below 100 K (Fig. 3b,d) correspond mostly to the R2 relaxations. In case of the $\boldsymbol{E} \perp c$, also weak $\varepsilon''(T)$ maxima could be distinguished above 100 K which are attributed to R1. Both $\varepsilon''(f)$ and $\varepsilon''(T)$ maxima are diffuse, therefore temperature dependences of the measured relaxation frequencies can be better defined from the temperature-frequency maps (Figs. 4,5). Both relaxation frequencies R1 and R2 at low temperatures follow the Arrhenius law:

$$f_R(T) = f_\infty e^{-\Delta E_T/T} = f_\infty e^{-\Delta E_{eV}/kT} \quad (4)$$

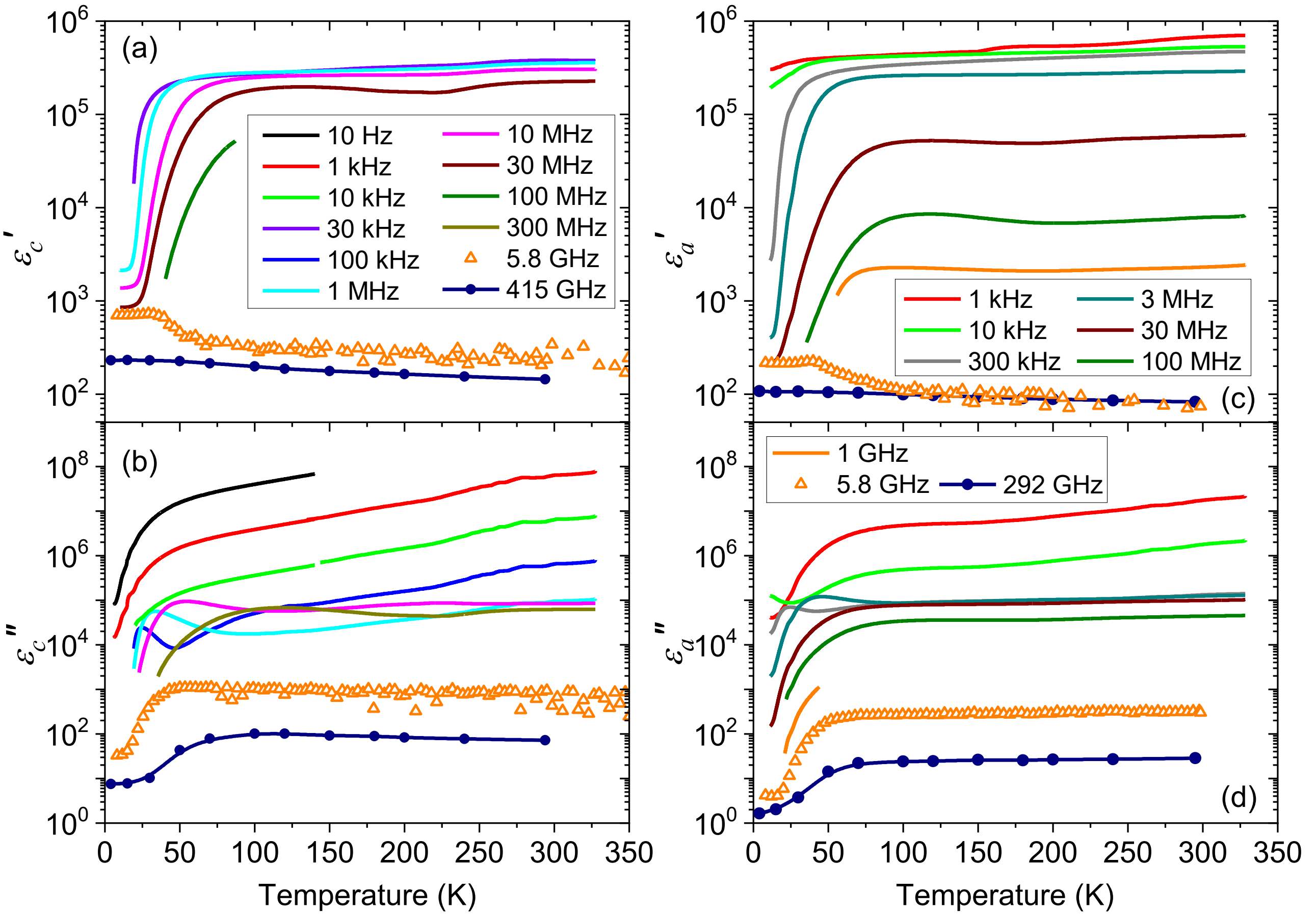

Fig. 3. Temperature dependences of the dielectric parameters of NTO at selected frequencies for $c$ and $a$ polarizations: (a,c) permittivity, (b,d) loss. MW data at 5.8 GHz are normalized to the more accurate HF (~100 MHz) and THz measurements at room temperature. Although the temperature dependence of the MW data is considered reliable, their absolute values were not used in the fitting shown in Fig. 1.

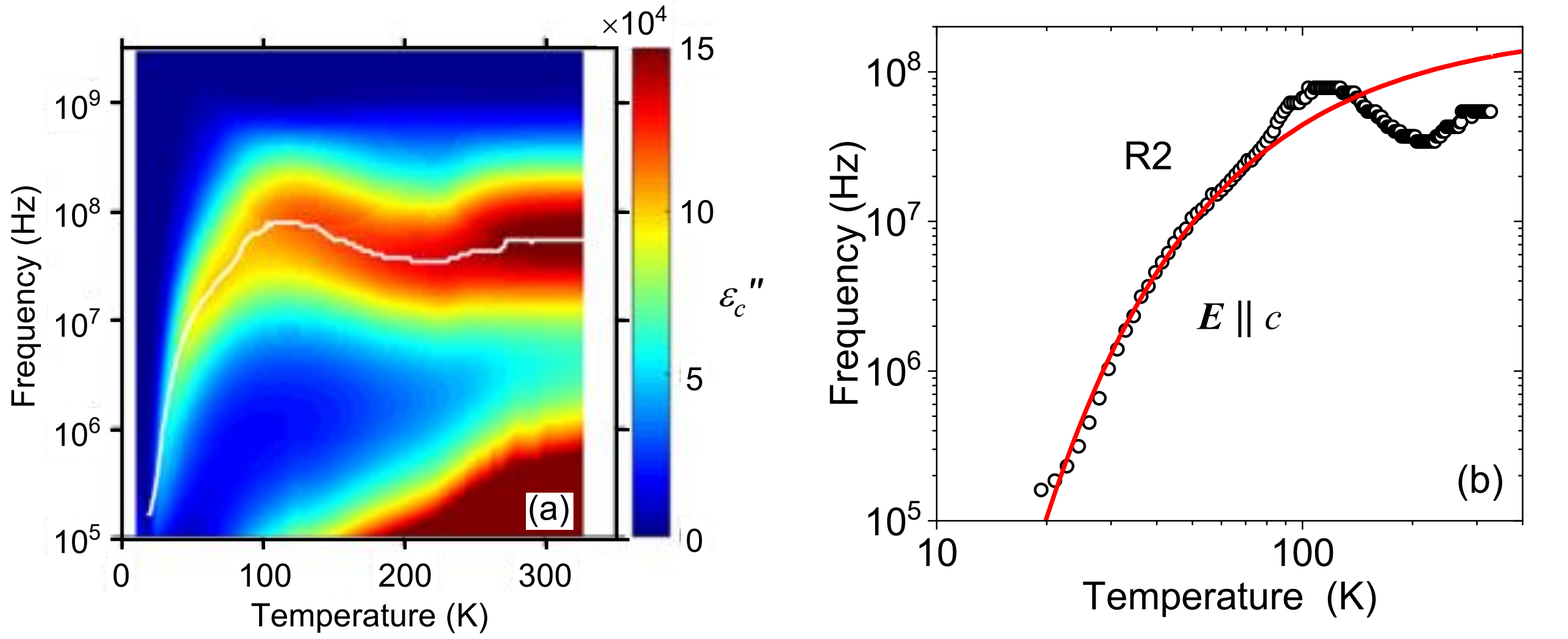

Fig. 4. R2 relaxation for the $\boldsymbol{E} \parallel c$ polarization: (a) temperature-frequency map of the dielectric loss, (b) temperature dependence of the mean frequency derived from the $\varepsilon''(f)$ maxima. The white line in (a) corresponds to the mean relaxation frequency. In (b), the experimental data of the $\varepsilon''(f)$ maxima are presented as symbols, the line corresponds to the Arrhenius fit (Eq. 4).

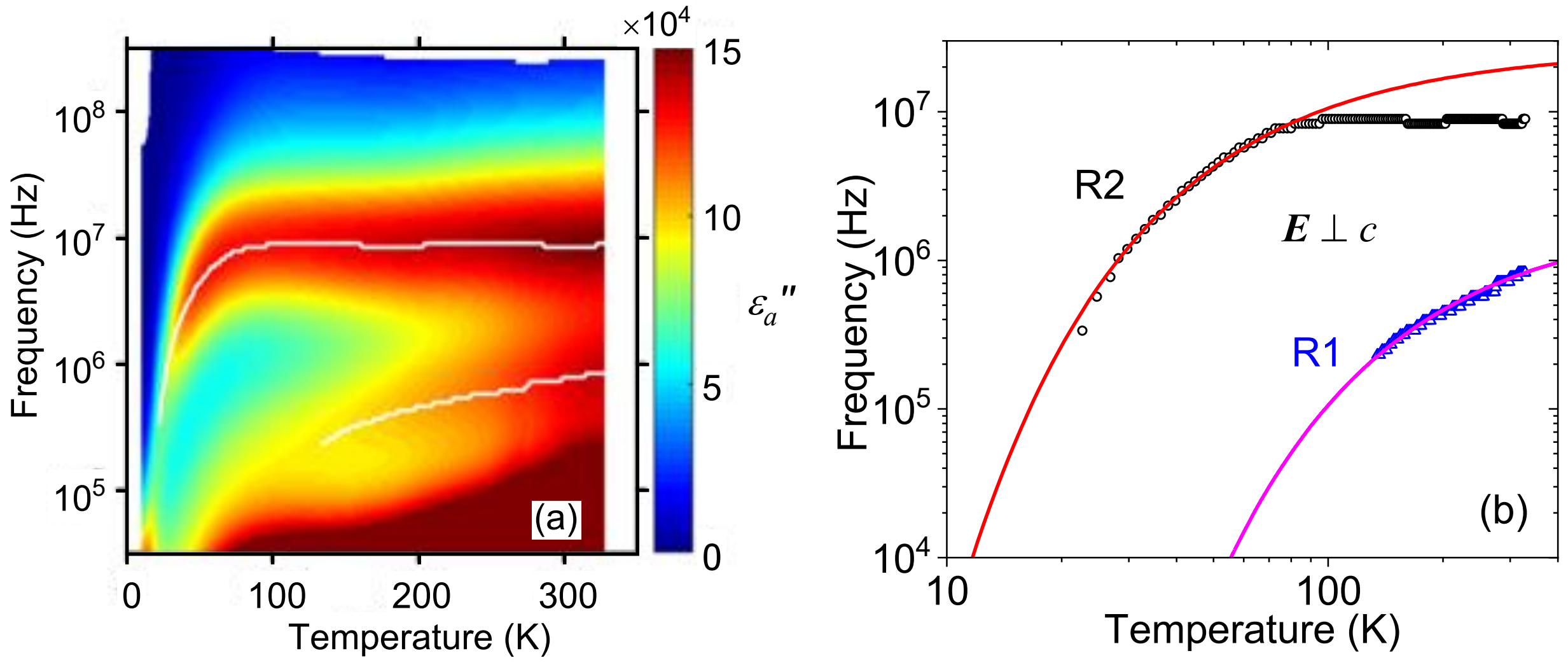

Fig. 5. R1 and R2 relaxations for the $\boldsymbol{E} \perp c$ polarization $a$: (a) temperature-frequency map of the dielectric loss, (b) temperature dependences of the $\varepsilon''(f)$ maxima. The white lines in (a) correspond to the mean relaxation frequencies. In (b), the experimental data of the $\varepsilon''(f)$ maxima are presented as symbols, the lines correspond to the Arrhenius fits (Eq. 4).

where $f_R(T)$ is relaxation frequency, $f_\infty$ is the relaxation frequency at $T \to \infty$, $\Delta E_T$ and $\Delta E_{eV}$ are activation energies in K and eV, respectively. The fit parameters are given in Table III.

Table III. Activation energies of the fitted relaxations. The third column specifies the fitting temperature range. NITO-001 and NTTO-001 correspond to $(Nb_{0.5}In_{0.5})_{0.01}Ti_{0.99}O_2$ and $(Nb_{0.5}Ti_{0.5})_{0.01}Ti_{0.99}O_2$ crystals of Ref. [31], respectively.

| AC field | Relaxation | $T$ (K) | $f_\infty$ (MHz) | $\Delta E_T$ (K) | $\Delta E_{eV}$ (meV) |
|---|---|---|---|---|---|
| $\boldsymbol{E} \parallel c$ | R2 | 20 – 80 | 200 | 151 | 13 |
| $\boldsymbol{E} \perp c$ | R2 | 15 – 80 | 26.5 | 92 | 7.9 |
| $\boldsymbol{E} \perp c$ | R1 | 139 – 330 | 2 | 295 | 25.4 |
| $\boldsymbol{E} \parallel c$ [31] | R in NITO-001 | 19 – 33 | $7.2\times10^3$ | 302 | 26 |
| $\boldsymbol{E} \parallel c$ [31] | R in NTTO-001 | 13 – 25 | $1.5\times10^3$ | 186 | 16 |

Considering the CP in NTO crystal and its frequency dispersion mainly as the result of the inhomogeneous conductivity with DL and some defects in the bulk sample, the higher-frequency relaxation R2 with lower $\Delta E$ could be attributed to the SBLC effect, as discussed later. Kakimoto et al. [31] also have observed Arrhenius-type relaxations in their NITO and NTTO single crystals (both containing ~1% Nb) as the $\varepsilon''(T)$ maxima in the $10^3 - 10^6$ Hz range at low temperatures, but measured only for $\boldsymbol{E} \parallel c$, with $\Delta E = 26$ and 16 meV, respectively, see Table III. The lower-frequency R1 could be attributed to electron hopping between $Ti^{3+}$ and $Ti^{4+}$ ions (free polaron). This assignment is supported by the significant $Ti^{3+}$ concentration detected by XPS and by the similar activation energy reported for a low-temperature relaxation attributed to $Ti^{3+}/Ti^{4+}$ hopping in undoped rutile ceramics [42]. Two segments can be distinguished in the temperature dependence of R2: the Arrhenius law is valid for $f_{R2}(T)$ below 80 K, but above 100 K it is nearly temperature independent (Figs. 4,5). The $f_{R1}(T)$ for $\boldsymbol{E} \perp c$ follows the Arrhenius law at high temperatures, but at lower temperatures it is overlapped by the DC conductivity; for $\boldsymbol{E} \parallel c$ the R1 relaxation is fully overlapped by the conductivity.

### C. LF dielectric spectra below 10 K

To reveal fully the freezing process of the relaxations, we measured the LF dielectric response down to 0.3 K. Two samples with different thicknesses but identical Ag-paste electrodes were measured for each polarization. Since the data were substantially independent of the sample thickness, in Fig. 6 we show them for the thinner sample for both polarizations. However, at low frequencies a remarkable difference was observed between the dielectric data with Ag-paste and sputtered Au electrodes, as shown in Fig. 6. Even at room temperature we observed similar large permittivity differences between the same samples with different electrodes as those at 10 K. However, at higher frequencies these differences vanish, as expected. Let us mention that large differences between the LF spectra with similar Ag and Au electrodes were previously observed also in our NITO ceramics in Ref. [24]. Therefore, the low-frequency difference at 10 K in Fig. 6 can be attributed to the SBLC effect which differs for both electrodes at LF, see Ref. [5] for more discussion.

In addition to our standard fit at 10 K from Fig. 1, in Fig. 6 we have shown also a fit without the CM contribution, using more general phenomenological Havriliak - Negami relaxations:

$$\varepsilon^*(f) = \varepsilon^*_{IR}(f) + \sum_j \frac{\Delta\varepsilon_{Rj}}{\left[1 + \left({if}/{f_{Rj}}\right)^{1-\alpha_j}\right]^{\beta_j}} + \frac{\sigma_0}{\varepsilon_0 f(i - f/\Gamma)} \quad (5)$$

with the corresponding parameters listed in Table IV. It is seen that such a fit cannot properly fit the HF and THz data, particularly for the $\boldsymbol{E} \perp c$ response. Therefore, we are convinced that the CM excitation is present in our NTO samples at all temperatures for both polarizations, even though the fits including our MW data are not perfect. The CM contribution also explains the higher LF permittivity at 0.3 K compared with its THz value, by ~230 and ~40 for $\boldsymbol{E} \parallel c$ and $\boldsymbol{E} \perp c$ polarization, respectively. Here, we reasonably assume that the THz permittivity at 10 K remains essentially unchanged on further cooling. Namely, as clearly seen in Fig. 6, at 0.3 K all the relaxations, including the SBLC effect, are frozen and therefore do not contribute to LF permittivity.

Table IV. Parameters of the Havriliak - Negami relaxations for both polarizations at 10 K, while the other parameters in Eq. 5 are kept unchanged.

| Polarization | $f_{Rj}$ (Hz) | $\alpha_j$ | $\beta_j$ | $\Delta\varepsilon_{Rj}$ |
|---|---|---|---|---|
| $\boldsymbol{E} \parallel c$ | 233 | 0 | 0.5 | $1.5\times10^5$ |
| | $4.3\times10^3$ | 0.24 | 1 | $7.8\times10^4$ |
| $\boldsymbol{E} \perp c$ | 0.013 | 0 | 0.56 | $1\times10^7$ |
| | $1.57\times10^4$ | 0.04 | 1 | $2.9\times10^5$ |

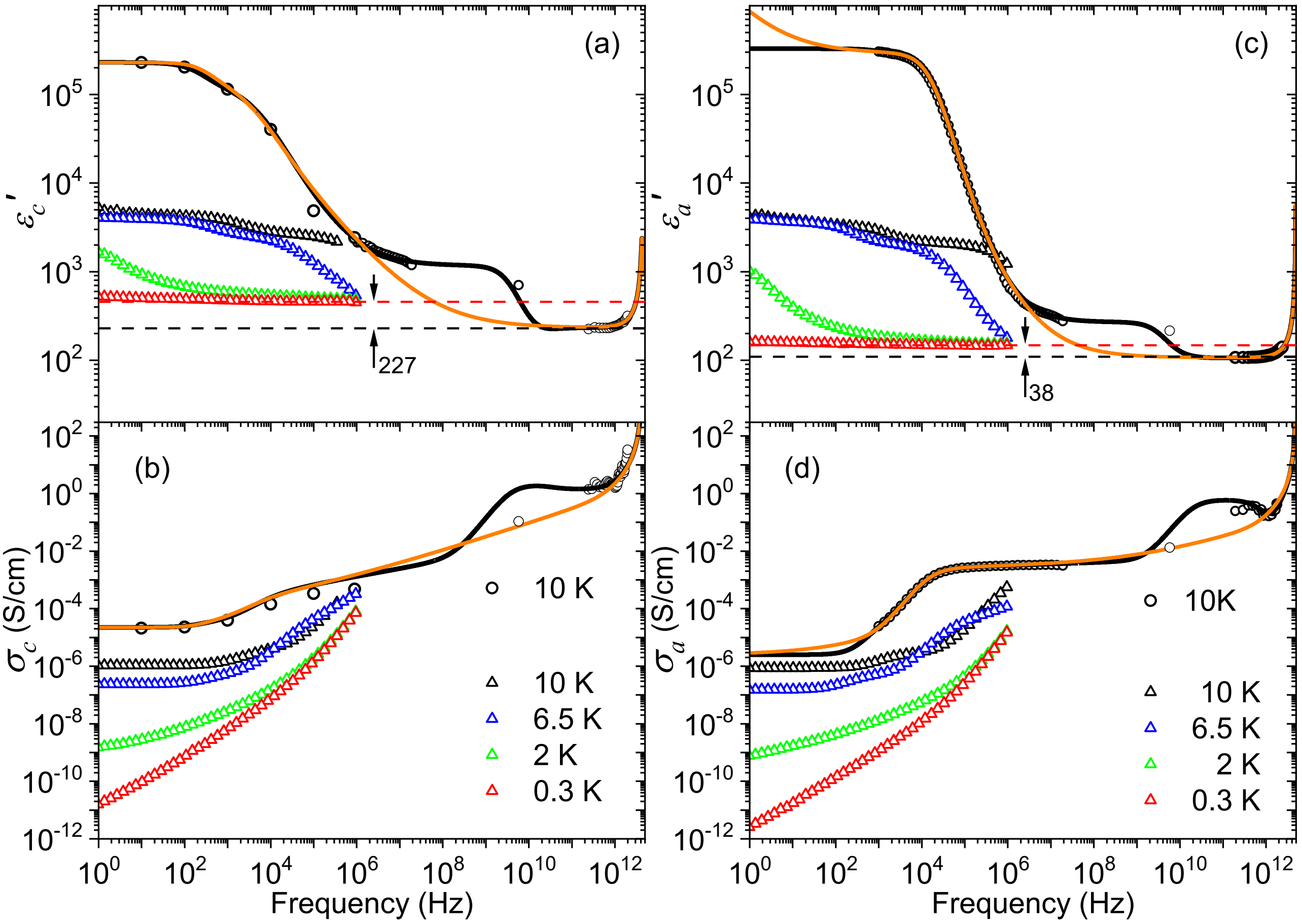

Fig. 6. (a,c) Broadband dielectric permittivity, (b,d) AC conductivity spectra of NTO for both polarizations at low temperatures (≤ 10 K), (a,b) $\boldsymbol{E} \parallel c$ polarization, LF spectra of the 540 μm thick sample, (c,d) $\boldsymbol{E} \perp c$ polarization, LF spectra of the 400 μm thick sample, both with Ag electrodes (triangles). At 10 K the data (up to the THz range) are compared with the results of samples with Au electrodes from Fig. 1 (circles). Black lines are the fits from Fig. 1. Orange lines represent the fits using Eq. 5. Low-temperature LF and THz permittivities, along with their differences, which account for the CM contributions, are shown by the dashed lines.

### D. IR reflectivity

In Fig. 7 we present the IR reflectivity spectra of NTO crystal for both polarization with their fits using Eq. 2 at three selected temperatures. The corresponding fitting parameters are shown in Tables I and II. The spectra in both polarizations are similar to those of undoped rutile crystal [17], see also Fig. 10, which shows the measured dielectric response in the THz range, along with the dielectric response calculated from fitting the THz data and IR reflectivity using Eq. 2. Comparison with our IR reflectivity spectra of undoped crystal indicates that the Nb-doping does not change the phonon mode frequencies. The main difference (slightly lower reflectivity maxima), results from a small increase in the phonon damping. We fitted the $A_{2u}$ ($\boldsymbol{E} \parallel c$) and $E_u$ ($\boldsymbol{E} \perp c$) reflectivity spectra using two and four TO modes, respectively. The SM for both polarizations corresponds to displacements of the Ti ions against O ions along and perpendicular to the $c$-axis [43]. By cooling from 300 to 10 K, the lowest-frequency polar modes soften by 33 and 22 cm$^{-1}$ in the $A_{2u}$ and $E_u$ spectra, respectively. Their dielectric strengths increase on cooling, while their oscillator strengths remain nearly constant, indicating that the SMs are not strongly coupled either to the other TO modes or to the CM. The small dip in the $\boldsymbol{E} \parallel c$ reflectivity spectra in the range of 300–500 cm$^{-1}$ is likely a polarization leakage. All the reflectivity minima sharpen on cooling due to decreasing phonon damping.

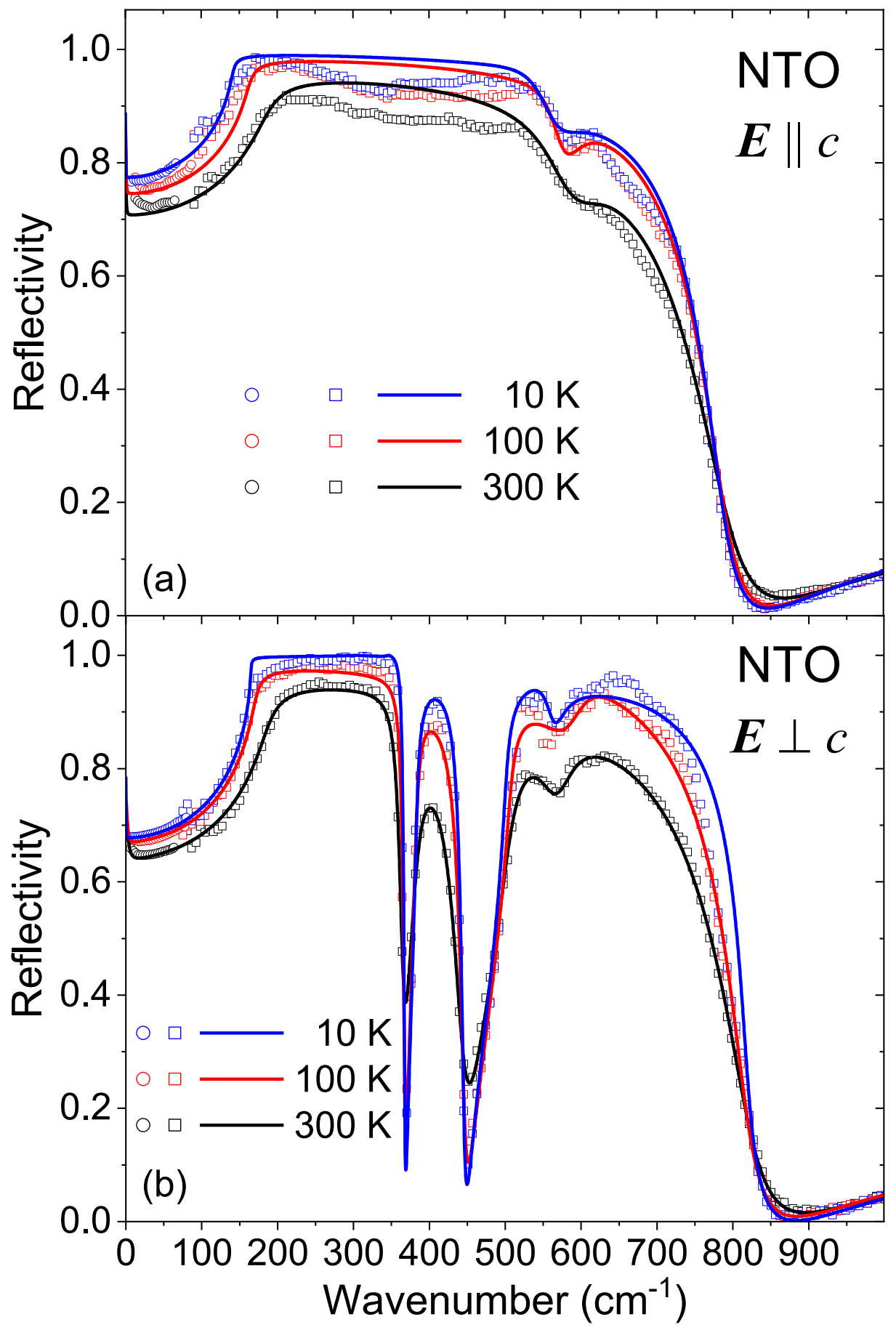

Fig. 7. IR reflectivity of the NTO crystal (squares) along with fits using Eq. 2 (lines), and THz reflectivity (circles) calculated from the measured dielectric response for both polarizations at three selected temperatures.

## IV. DISCUSSION

### A. Activation energies from *ε''(f, T)* maxima and *σ(f)* plateaus

From our data it is seen that the activation energies of relaxation frequencies vary quite appreciably for both polarizations, temperature range and type of electrodes. The higher-frequency relaxation R2 can be assigned to the SBLC effect, see later our modelling of the DL and also previous modelling of the NITO ceramics [26] and generally of the core-shell structures [44]. The SBLC effect also accounts for the two AC conductivity plateaus seen in our spectra in Fig. 1. The LF one represents the effective DC conductivity, which accounts for the influence of the low-conductivity DL in series with the much higher bulk sample conductivity. The HF conductivity plateau represents quite well the bulk conductivity of the sample since it stays frequency independent within this frequency range (as supported by our Drude model). In this frequency range, the influence of DL can be neglected due to its negligible thickness compared to the sample thickness. To check it, in Sec. **IV B** we will analyze in more detail the temperature dependence of our NTO conductivity for $\boldsymbol{E} \perp c$, for which we have measured also the 4-point DC conductivity data ($\sigma_{DC}$) from 300 down to 2 K.

In Fig. 8, the temperature dependent $\sigma_{DC}$ is presented together with the LF ($\sigma_{1\,kHz}$) and HF ($\sigma_{100\,MHz}$) AC conductivities corresponding roughly to the $\sigma(f)$ plateaus in the kHz and 100 MHz ranges, respectively (see Fig. 1f). The values of $\sigma_{DC}$ and HF AC conductivity $\sigma_{100\,MHz}$ are very close and their temperature dependences obey the Arrhenius law between 10 and 25 K with the same activation energy of 11 meV (Table V):

$$\sigma = \sigma_{\infty} e^{-\Delta E_T/T} = \sigma_{\infty} e^{-\Delta E_{eV}/kT} \tag{6}$$

where $\sigma_\infty$ is the conductivity for $T \rightarrow \infty$.

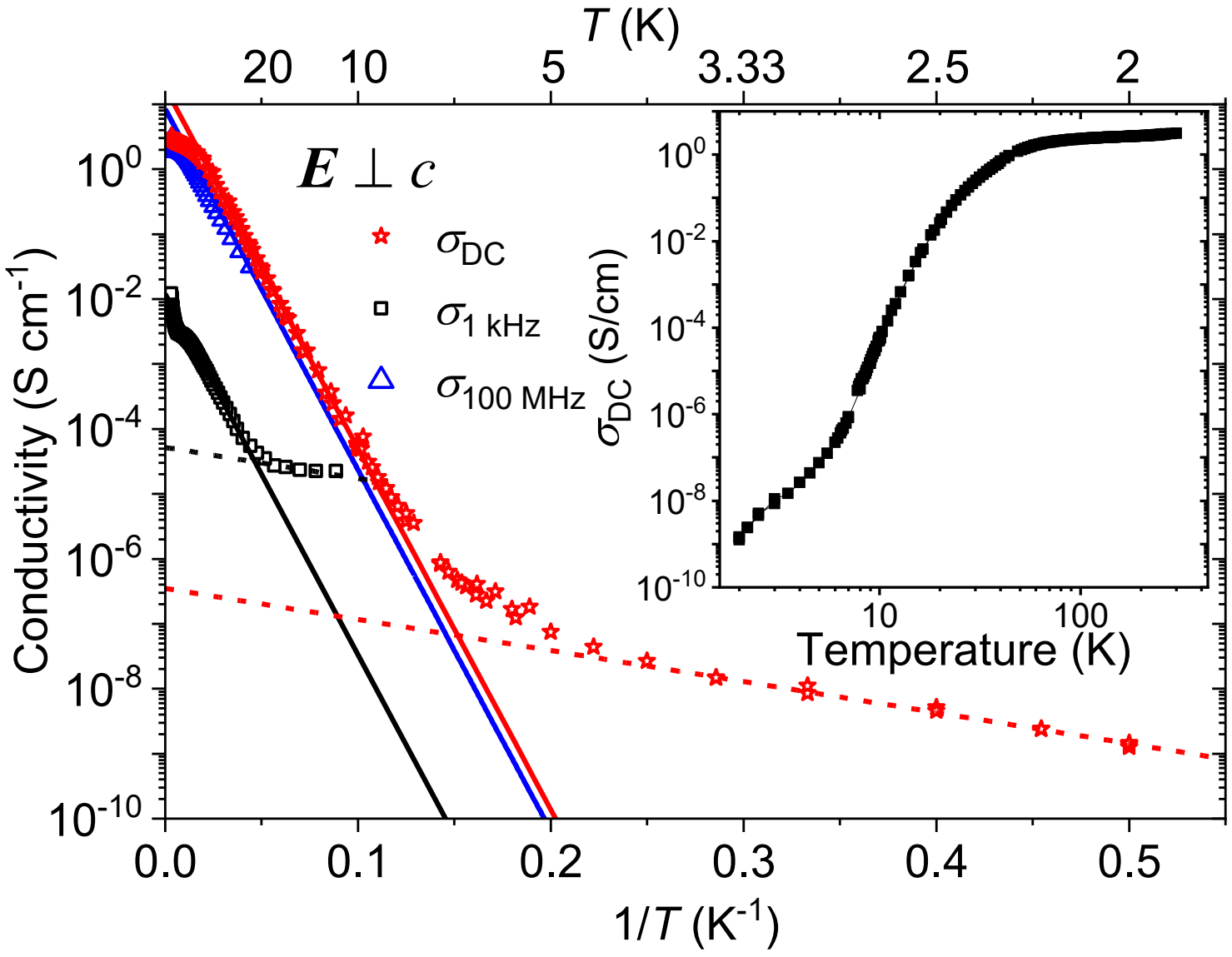

Fig. 8. Arrhenius plot of the temperature dependence of 4-point DC conductivity ($\sigma_{DC}$), LF AC conductivity ($\sigma_{1\ kHz}$) and HF AC conductivity ($\sigma_{100\ MHz}$) for $\boldsymbol{E} \perp c$. Experimental data are presented by the symbols, lines correspond to the Arrhenius fits (Eq. 6). Inset: Temperature dependence of the bulk DC conductivity measured by the 4-point method.

Table V. Activation energy of the DC, LF and HF conductivities for $\boldsymbol{E} \perp c$. The second column specifies the fitting temperature range.

| | Temperature (K) | $\sigma_\infty$ (S/cm) | $\Delta E_T$ (K) | $\Delta E_{eV}$ (meV) |
|---|---|---|---|---|
| $\sigma_{DC}$ | 10 – 50 | 18 | 128 | 11 |
| | 2 – 4 | 3.5×10$^{-7}$ | 11 | 1 |
| $\sigma_{1\ kHz}$ | 25 – 100 | 0.012 | 128 | 11 |
| | 12.5 – 20 | 5.2×10$^{-5}$ | 11 | 1 |
| $\sigma_{100\ MHz}$ | 25 – 80 | 8.5 | 128 | 11 |

Above 25 K, the values of $\sigma_{1\ kHz}$ are much smaller than $\sigma_{DC}$ and $\sigma_{100\ MHz}$ but follow the Arrhenius law with the same activation energy of $\Delta E = 11$ meV which corresponds to the high-temperature DC-conductivity mechanism. The low-temperature conductivity is characterized by $\Delta E$ = 1 meV and corresponds to $\sigma_{DC}$ and $\sigma_{1\ kHz}$ conductivity below 4 and 12.5 K, respectively. However, the $\sigma_{1\ kHz}$ conductivity does not represent properly the LF conductivity plateau below ~30 K, which should appear at lower frequencies, see Fig. 1. Therefore, the agreement of the $\sigma_{DC}$ activation energy below 4 K with that of the $\sigma_{1\ kHz}$ conductivity below 12.5 K appears to be rather accidental. Similar values, $\Delta E$ ~2 meV and ~11 meV in the 50-100 K range were also reported for the strengths of two broad overlapping middle IR $\boldsymbol{E} \perp c$ absorption bands in the Nb-doped rutile crystals (stronger one at ~ 6500 cm$^{-1}$ and weaker one at ~3100 cm$^{-1}$) by Weiser et al. [45] (see their Fig. 7), ascribed to the effect of small polarons. Below ~50 K (measured down to 15 K) both their strengths remain non-zero and essentially temperature independent. However, their contributions to IR dielectric losses are too small to be detected by our IR reflectivity measurements.

The 4-points conductivity experiment eliminates the influence of the DL and provides the bulk DC conductivity. Due to the relaxation dispersion, the AC conductivity increases with frequency, and the DL gradually loses its effect on the measured AC conductivity, so that the $\sigma_{100\ MHz}$ corresponds nearly to the DC conductivity of the bulk NTO. Let us mention that the DC conductivity of bulk NTO ceramics as well as single crystals by using the 4-point technique was already measured and discussed by Itakura et al. [46] and at high temperatures by Baumard and Tani [47]. Our results are compatible with the previous studies.

## B. SBLC modelling of the dielectric spectra

For the modelling we chose the $\boldsymbol{E} \perp c$ dielectric spectra and used the corresponding 4-points DC-conductivity measurements. We assume that the CP can be essentially explained by the formation of low-conductivity DL at the sample-electrode interfaces. A similar approach was used to describe the CP in co-doped $TiO_2$ ceramics, where an additional contribution from low-conductivity grain boundaries was also considered [26]. Of course, in the present case we deal with a single crystal where the grain-boundary effect is absent. For consideration of the DL, it allows us to use a simple series plate-capacitor model:

$$\varepsilon_{eff}^{-1} = x_{bulk}\varepsilon_{bulk}^{-1} + (1 - x_{bulk})\varepsilon_{DL}^{-1} \tag{7}$$

This equation can be equivalently described by the Lichtenecker model [48] for $\alpha = -1$, corresponding to the maximum depolarizing field with no percolation of any component, regardless of their concentration. Effective dielectric response $\varepsilon_{eff}$ in Eq. 7 is fitted by using two fitting parameters: the DL thickness (volume concentration $x_{DL} = 1 - x_{bulk}$) and its conductivity ($\sigma_{DL}$). The constant permittivity value ($\varepsilon_{bulk}^{GHz}$) was taken from fits shown in Fig. 1d in the GHz range including the CM contribution. The same permittivity value was used for both the bulk and DL ($\varepsilon^{GHz} = \varepsilon_{bulk}^{GHz} = \varepsilon_{DL}^{GHz}$). The almost frequency-independent bulk conductivity up to the THz range was described by the Drude term with $\Gamma$=100 THz as used in Eq. 1:

$$\varepsilon_{bulk} = \varepsilon^{GHz} + \frac{\sigma_{DC}}{\varepsilon_0 f(i - f/\Gamma)}, \qquad \varepsilon_{DL} = \varepsilon^{GHz} + \frac{\sigma_{DL}}{\varepsilon_0 f(i - f/\Gamma)} \tag{8}$$

The bulk DC conductivity $\sigma_{DC}$ was determined using the 4-points DC-conductivity measurement. The input and fitting parameters are summarized in Table VI, and the corresponding fits are shown in Fig. 9.

Table VI. Modelling parameters of the dielectric spectra of our NTO sample (thickness 1.3 mm) for $\boldsymbol{E} \perp c$, using Eqs. 7 and 8.

| Temperature | $\varepsilon^{GHz}$ | $x_{DL}$ | $\sigma_{DC}$ (S/cm) | $\sigma_{DL}$ (S/cm) | DL thickness (nm) |
|---|---|---|---|---|---|
| 30 K | 184 | $5\times10^{-4}$ | 0.25 | $8.8\times10^{-8}$ | 324±15 |
| 100 K | 182 | $4.1\times10^{-4}$ | 2.3 | $1.2\times10^{-6}$ | 268±10 |
| 300 K | 163 | $2.4\times10^{-4}$ | 3 | $2.4\times10^{-6}$ | 158±7 |

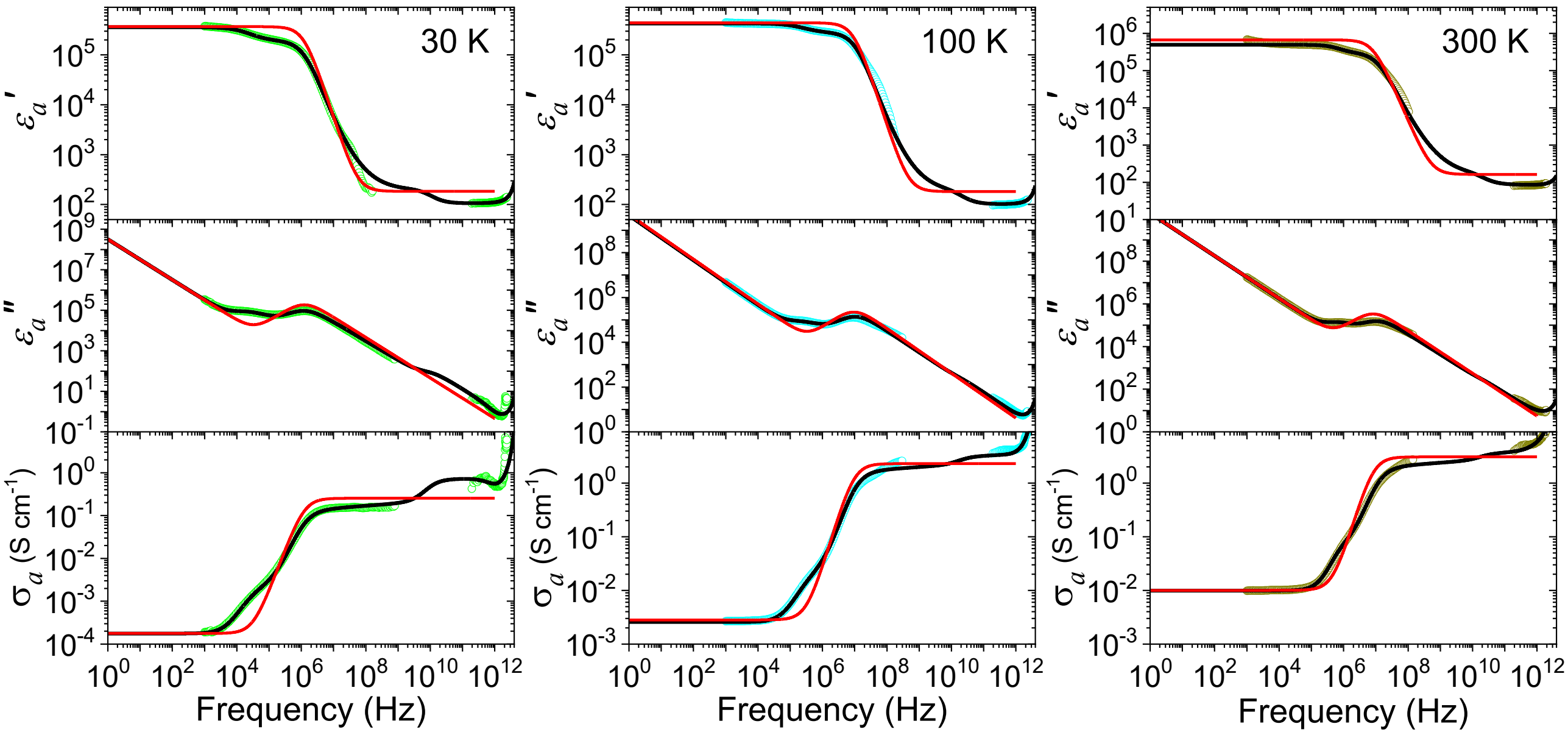

Fig. 9. Dielectric spectra (symbols) for $\boldsymbol{E} \perp c$ at 30, 100 and 300 K are shown with fits (black lines) and modelling results (red lines) using Eqs. 1,2 and Eqs. 7,8, respectively.

From Fig. 9 it is seen that the main effect of the CP can be well explained by the existence of DL whose thickness slightly increases on cooling down to 30 K. It agrees well with the SBLC modelling on the NITO ceramics [24-26]. At even lower temperatures < 30 K, due to the strongly decreasing bulk conductivity, our fits deteriorate and the DL thickness unrealistically increases, indicating that the SBLC model loses its validity. It is evident that the SBLC effect is mainly represented by the R2 relaxation in the MHz range which slows down below ~80 K following the Arrhenius law (see Fig. 5 and Table III).

**C. Comparison of NTO with undoped rutile crystal**

The IR reflectivity spectra maxima of NTO in the 10–300 K temperature range are consistently slightly lower than those of undoped $TiO_2$ crystal at corresponding temperatures. This is attributed to slightly higher optical phonon damping, as determined from the fitting using Eq. 2, while the phonon mode frequencies remain essentially unaffected by the Nb doping. However, a remarkable difference is observed in the dielectric losses in THz range and below it in the HF-MW range, caused by the presence of the overdamped CM in NTO at all temperatures (see also the 5.8 GHz data in Fig. 3), which is completely absent in the undoped rutile. The fits for both crystals and polarizations, along with our experimental data and MW data for the undoped $TiO_2$ from the literature [49,50] at three selected temperatures, are shown in Fig. 10.

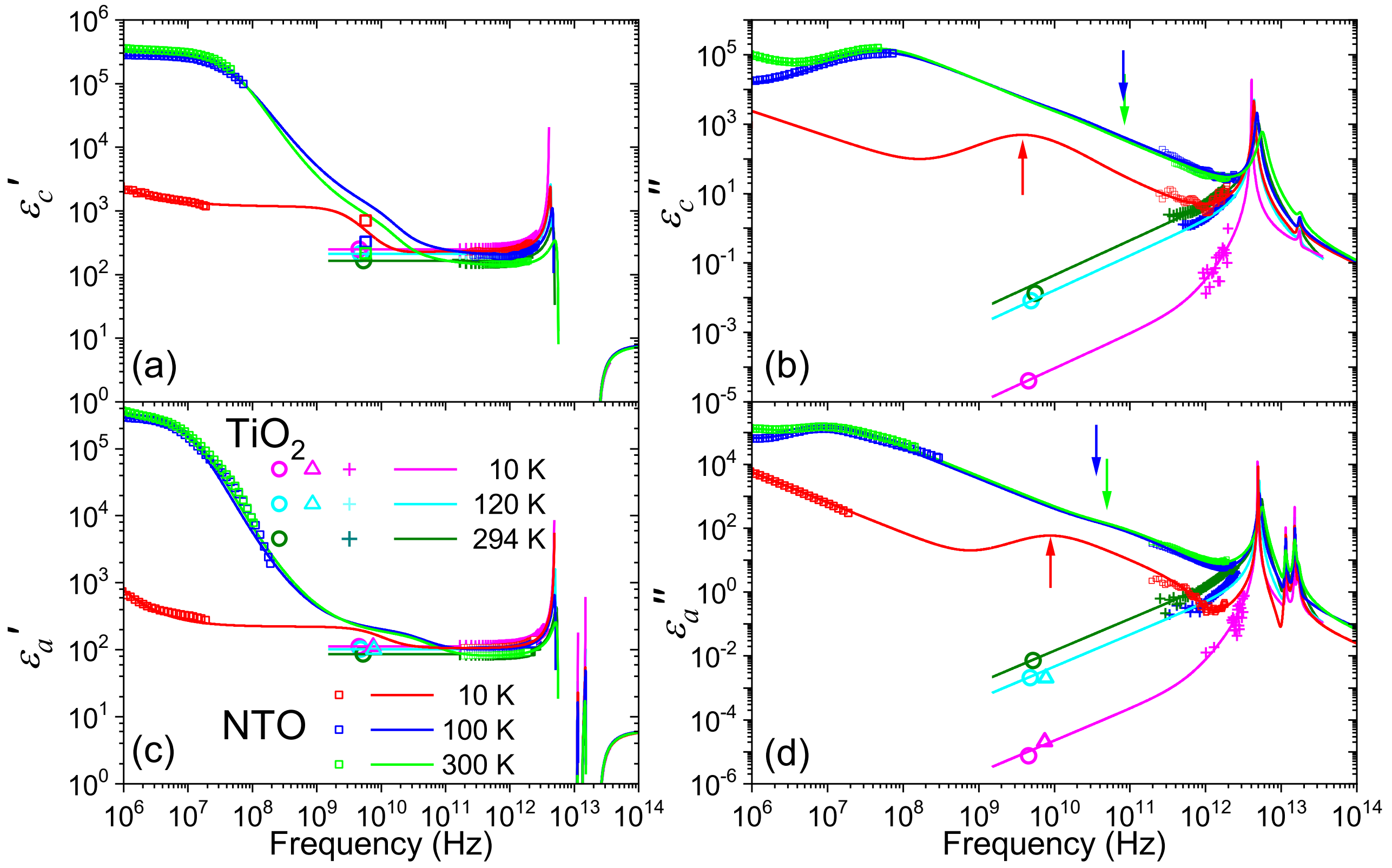

Fig. 10. Comparison of the dielectric spectra in the $10^6$-$10^{14}$ Hz range of the NTO with undoped $TiO_2$ crystal for both polarizations. Symbols are experimental data and lines are fits using Eqs. 1,2. MW data near 10 GHz for $TiO_2$ are taken from Tobar et al. [49] (circles) and Klein et al. [50] (triangles). Fits for the $TiO_2$ crystal were obtained using our THz-IR experimental data and Eq. 2. In (b,d) the CM frequencies from THz-IR fits are indicated by arrows.

From Tables I, II and Fig. 2 it is seen that the dielectric strength of the CM is rather strong, comparable or even stronger than that of the phonons, and on cooling it even shows tendency to strengthen. It explains why the low-temperature permittivity remains appreciably higher than that of undoped samples even below 10 K, where all the relaxations are frozen [2728,3031]. This is also in good agreement with the new data reported in Ref. [29] for NTO single crystals with Nb concentrations of up to 2%. The density functional theory (DFT) calculations in this reference (see their Fig. 8) show that even the dynamic properties of the molecular polaron in the 10-100 GHz range are in good agreement with our data, including their quantum tunneling at low temperatures, clearly seen in our data, and increasing CM frequency with temperature. Let us also recall that the mid-IR absorption bands in Nb-doped rutile single crystals, ascribed to small polaron effects, remain active down to the liquid-He temperatures [45]. They are known to co-exist with the delocalized electrons which yield the Drude-type conductivity [45,51] determined by our 4-point measurements. More quantitative properties concerning the first-time detected CM in NTO crystal would require more detailed MW measurements.

## V. CONCLUSIONS

Our broadband dielectric measurements of the NTO crystal up to the mid-IR range provide clear evidence that the main reason of the CP in the Nb-doped rutile is due to a near-electrode depletion layer (DL) of ~100-300 nm thickness with the room-temperature conductivity about six-orders of magnitude lower than that of the bulk. Both bulk DC and effective AC conductivities are thermally activated at low temperatures with small activation energies of ~1 meV, which increase by one-order of magnitude at higher temperatures and above ~80 K both conductivities become nearly temperature independent. The activation energies roughly correspond to those of two mid-IR absorption bands associated with small polarons [45]. In the microwave range, we have observed a new overdamped excitation in the doped rutile crystal, which is not thermally activated and explains the higher low-temperature LF permittivity in previous NITO measurements compared to undoped rutile crystal [27,28,30,31] and to the THz measurements of our NTO. Its origin can be attributed to small polarons, as recently proposed by DFT calculations in Ref. [29]. On the other hand, the Nb-doping causes only a slight increase in the damping of optical phonons, without changing their frequencies.

## ACKNOWLEDGMENTS

The authors thank M. Setvín from the Faculty of Mathematics and Physics, Charles University Prague, for fruitful discussions on polarons and M. Trousilová for the X-ray diffraction sample orientations. The work was supported by the Czech Science Foundation (Project No. 24-10791S) and by the project TERAFIT - CZ.02.01.01/00/22_008/0004594 co-financed by the European Union and the Ministry of Education, Youth and Sports of the Czech Republic.

# Title: Supplementary Material

# Description: Additional details on the XPS experiment and data analysis.

## Giant dielectric permittivity in Nb-doped rutile crystals

D. Nuzhnyy,[1] V. Bovtun,[1] J. Petzelt,[1] M. Savinov,[1] M. Kempa,[1] P. Levinský,[2] K. Beranová,[2] P. Vaněk,[1] T. Kmječ,[1] T. Ostapchuk,[1] P. Kužel,[1] J. Hlinka,[1] D. Crandles,[3] M. Cosco,[3] Y. Hashimoto,[4] H. Taniguchi,[4] S. Kamba[1]

[1]*Institute of Physics of the Czech Academy of Sciences, Na Slovance 2, 182 00 Prague 8, Czechia*
[2]*Institute of Physics of the Czech Academy of Sciences, Cukrovarnická 10/112, 162 00 Prague 6, Czechia*
[3]*Department of Physics, Brock University, St. Catharines, Ontario L2S 3A1, Canada*
[4]*Department of Physics, Nagoya University, Nagoya 464-8602, Japan*

**XPS experimental details**

XPS measurements were performed using the AXIS Supra photoelectron spectrometer (Kratos Analytical). Monochromatized Al Kα was used for all measurements while the emission current and the pass energy of the analyzer were 5 mA - 80 eV, 10 mA – 40 eV and 15 mA – 10 eV for survey scans, low-energy-resolution regions (Zn 2p, O 1s, Ti 2p, Ca 2p, K 2p, C 1s, Nb 3d, and Ti LMM) and high-energy-resolution regions for detailed analysis (Ti 2p, O 1s, Nb 3d, C 1s), respectively. The sample was insulated by an adhesive paper tape, and the charge compensation (low-energy electrons) was used during the whole measurement.

The XPS spectra were analyzed in the ESCApe software (Kratos Analytical, ver. 1.7.0.0925), using its built-in library of sensitivity factors to calculate the atomic concentration of elements. All spectra were calibrated to 284.8 eV of the C-C peak in the C 1s region. Spectra were fitted using Shirley background and G*L peaks with 0.3 ratio, except for the $Ti^{4+}$ $2p_{3/2}$ component where the G/L ratio was adjusted to 0.58.

**Data analysis**

XPS measurements indicate that titanium is present in the rutile structure, as evidenced by the modified Auger parameter of 873.5 eV, which is in good agreement with the value reported by Ingo et al. [1]. The $Ti/O_{Lattice}$ ratio is 0.53. Approximately 3.2 at.% of titanium is in the $Ti^{3+}$ oxidation state [2]. Niobium accounts for approximately 1.6 at.% of cations in the oxide structure and is detected in a single chemical state, with a Nb $3d_{5/2}$ binding energy of 206.8 eV. This value lies between those reported for $Nb_2O_5$ (207.5 eV) and $NbO_2$ (206.2 eV) [3,4], but is typical for Nb in substitutional positions in rutile $TiO_2$ [3,4,5]. According to the recent work of Van An Dinh et al. [6], $Nb^{5+}$ doped in $TiO_2$ forms the $Nb^{5+}$ - $Ti^{3+}$ pairs while the released electron is localized between adjacent Nb and Ti atoms, in closer proximity to Ti. This results in forming a slightly more negative effective charge on the $Nb^{5+}$ dopant: $Nb^{x+}$ ($4 < x < 5$). Our observation of an intermediate Nb 3d binding energy is consistent with this interpretation and supports the formation of small polarons in Nb-doped $TiO_2$.

The formation of small polarons requires equal number of $Nb^{x+}$ and $Ti^{3+}$ ions. However, the atomic concentration of $Ti^{3+}$ ions is slightly higher in our $NbTiO_x$ crystal. Therefore, the surplus $Ti^{3+}$ ions may have another origin, such as crystal defects and oxygen vacancies. XPS cannot directly observe oxygen vacancies, but presence of $Ti^{3+}$ ions and oxygen-containing surface adsorbates (Table S1, Figure S4) indicate existence of such defects [2].

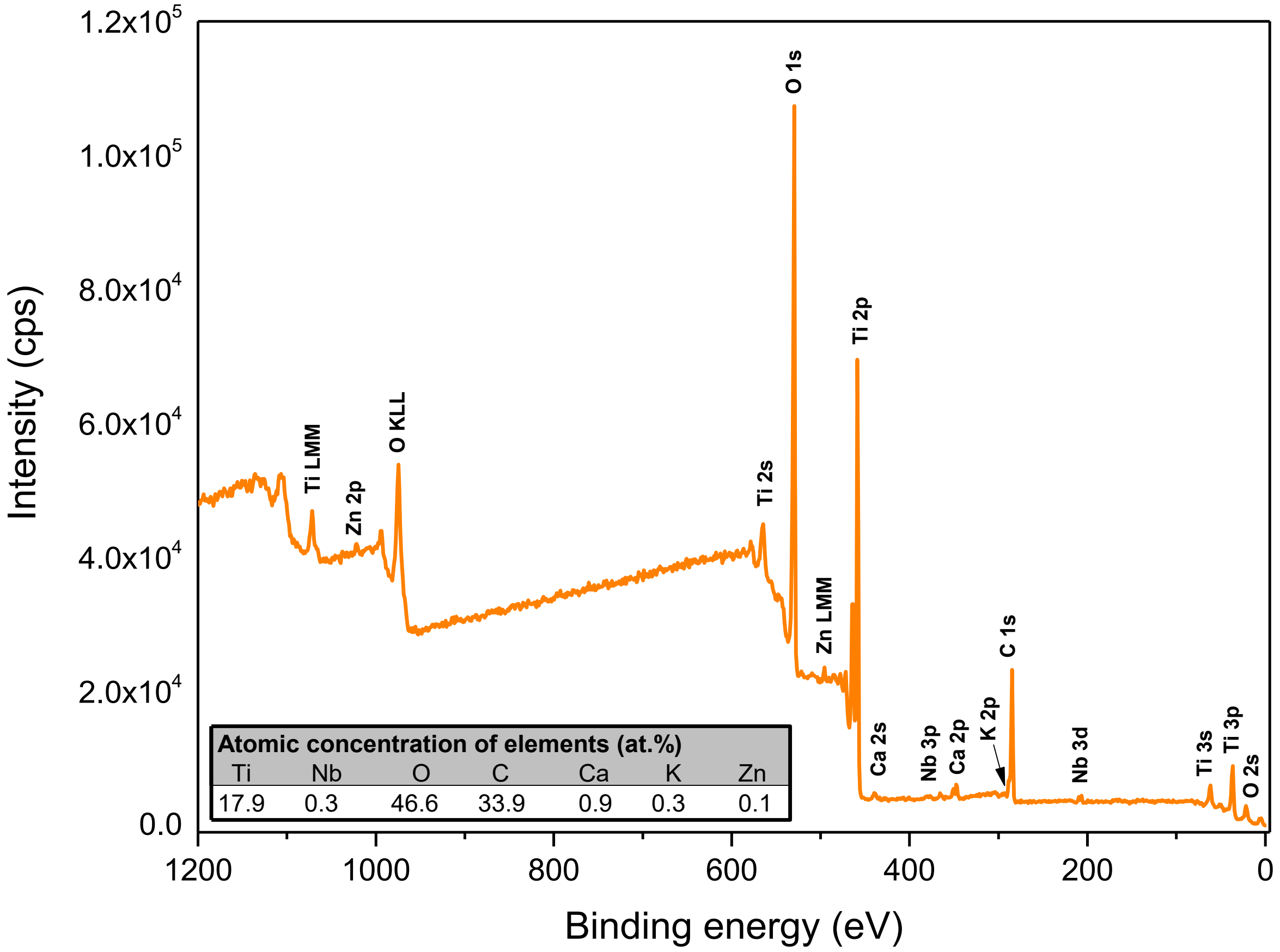

**Figure S1:** Survey XPS scan of $NbTiO_x$ with indicated photoelectron and Auger lines. Atomic concentration of elements was calculated from the Ti 2p, Nb 3d, O 1s, C 1s, Ca 2p, K 2p and Zn 2p core levels measured with emission 10 mA and pass energy of 40 eV. Carbon and traces of K, Ca, and Zn were detected as surface contaminants.

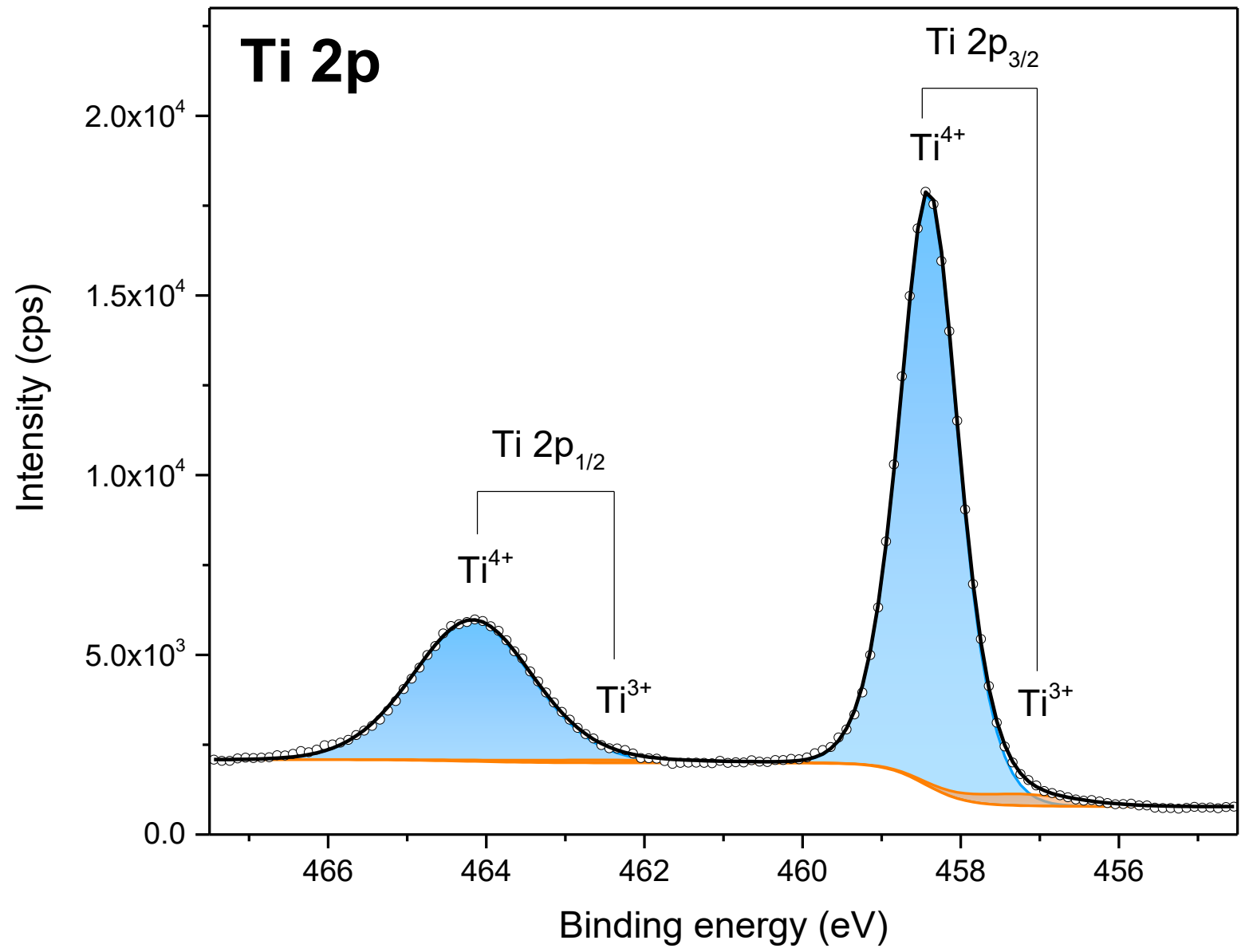

**Figure S2:** High-energy-resolution Ti 2p XPS spectrum consists of two spin-orbit doublets originating from $Ti^{4+}$ ($TiO_2$-rutile, blue) and $Ti^{3+}$ (orange) [2]. Approximately 3.2 at.% of titanium is in the $Ti^{3+}$ oxidation state.

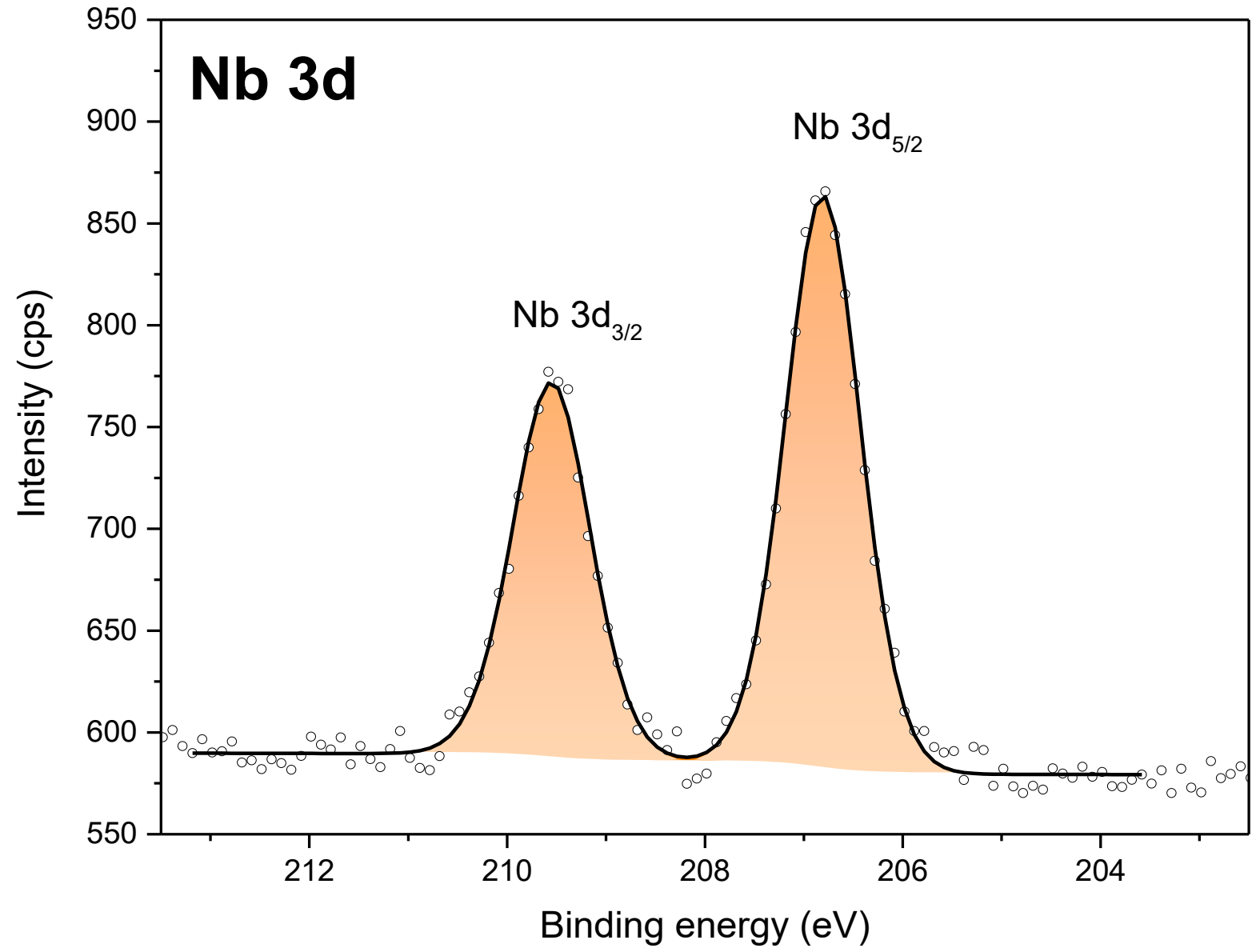

**Figure S3:** High-energy-resolution Nb 3d XPS spectrum consists of single spin-orbit doublet originating from $Nb^{5+}$ ions in the $TiO_2$ mattrix [3,4].

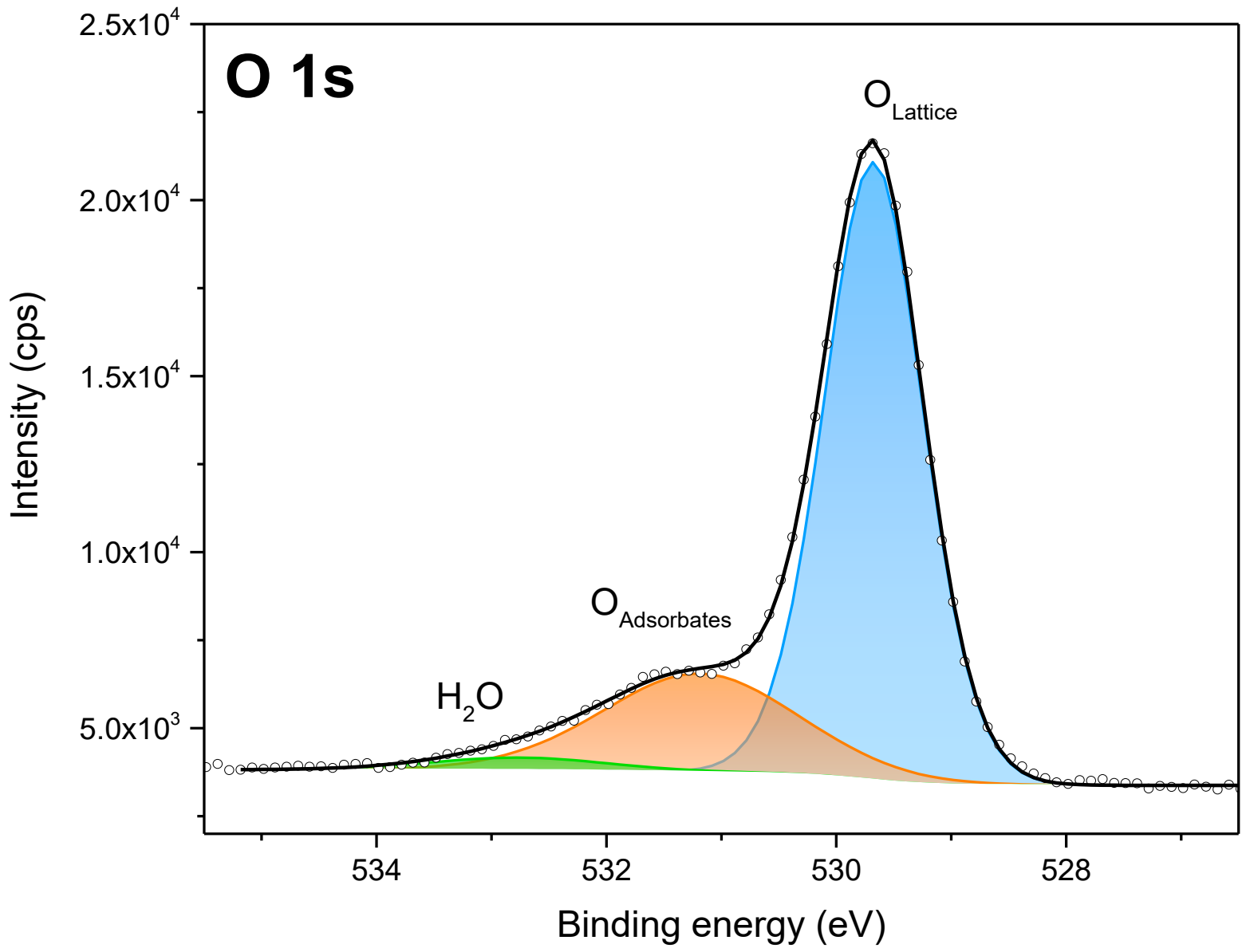

**Figure S4:** High-energy-resolution O 1s XPS spectrum consists of lattice oxygen (blue), oxygen-containing adsorbates in strong binding environment, such as -OH, -CO or -COOH (orange) and molecular water (green) [2,7].

**Table S1:** Fitting parameters of XPS core level regions: binding energy (BE), full-width-in-half-the-maximum (FWHM), atomic concentration of components derived from detailed analysis of high-energy-resolution XPS spectra measured with 15 mA and pass energy of 10 eV.

| Element | Photoelectron | BE (eV) | FWHM (eV) | Attribution | Atomic concentration (at.%) |
|---|---|---|---|---|---|
| Ti | Ti $2p_{3/2}$ | 458.4 | 0.9 | $Ti^{4+}$ (rutile) | 12.3 |
| | | 457.3 | 1.7 | $Ti^{3+}$ | 0.4 |
| | Ti $2p_{1/2}$ | 464.2 | 1.7 | $Ti^{4+}$ (rutile) | 5.6 |
| | | 462.5 | 3.4 | $Ti^{3+}$ | 0.2 |
| | Ti LMM | 1071.5 | | | |
| Nb | Nb $3d_{5/2}$ | 206.8 | 0.9 | $Nb^{x+}$-$TiO_2$ | 0.2 |
| | Nb $3d_{3/2}$ | 209.6 | 0.9 | $Nb^{x+}$-$TiO_2$ | 0.1 |
| O | O 1s | 529.7 | 1.0 | Lattice oxygen | 35.2 |
| | | 531.2 | 2.0 | Adsorbates | 11.0 |
| | | 532.8 | 1.8 | $H_2O$ | 1.3 |
| C | C 1s | 284.8 | 1.1 | C-C | 27.8 |
| | | 286.3 | 1.2 | C-O | 2.7 |
| | | 287.6 | 1.2 | C=O | 0.0 |
| | | 288.4 | 1.3 | O=C-O | 3.1 |